\documentclass[sigconf]{acmart}
\AtBeginDocument{}

\copyrightyear{2026}
\acmYear{2026}
\setcopyright{cc}
\setcctype{by}
\acmConference[PASC '26]{Platform for Advanced Scientific Computing Conference}{June 29-July 01, 2026}{Bern, Switzerland}
\acmBooktitle{Platform for Advanced Scientific Computing Conference (PASC '26), June 29-July 01, 2026, Bern, Switzerland}
\acmDOI{10.1145/3815572.3815749}
\acmISBN{979-8-4007-2734-4/2026/06}

\usepackage{xcolor}
\usepackage{cleveref}

\crefname{figure}{Fig.}{Figs.}  

\begin{document}

\title{The Memory Scaling of Reverse-Mode Differentiation in Particle Accelerator Simulations with Space Charge}

\author{Arjun Dhamrait}
\author{Edoardo Zoni}
\author{Axel Huebl}
\author{Ji Qiang}
\author{Chad Mitchell}
\affiliation{
  \institution{Lawrence Berkeley National Laboratory}
  \city{Berkeley}
  \state{California}
  \country{USA}
}

\author{Ryan Roussel}
\affiliation{
  \institution{SLAC National Laboratory}
  \city{Stanford}
  \state{California}
  \country{USA}
}

\author{Jan Kaiser}
\affiliation{
  \institution{Deutsches Elektronen-Synchrotron DESY}
  \city{Hamburg}
  \country{Germany}
}

\author{Chenran Xu}
\affiliation{
  \institution{Argonne National Laboratory}
  \city{Lemont}
  \state{Illinois}
  \country{USA}
}

\author{Jean-Luc Vay}
\author{Remi Lehe}
\affiliation{
  \institution{Lawrence Berkeley National Laboratory}
  \city{Berkeley}
  \state{California}
  \country{USA}
}

\renewcommand{\shortauthors}{Dhamrait et al.}

\begin{abstract}
The recent development of \emph{differentiable} simulation codes for particle accelerators has enabled gradient-based workflows that promise finer control and more realistic modeling of accelerator facilities. However, when using reverse-mode automatic differentiation, the memory usage continuously increases during the simulation, and can potentially exceed the available hardware memory -- especially when costly space charge computation is included. To study the memory requirements for differentiable simulations, we have implemented space charge in Cheetah, a PyTorch-based beam tracking code that supports reverse-mode differentiation. We find that the memory usage for reverse-mode differentiation grows linearly with the number of macroparticles and cells, and that it is proportional to the number of space charge kicks involved in the simulation. This general scaling can be used to evaluate whether a given differentiable simulation is feasible given hardware memory constraints.
\end{abstract}



\maketitle

\section{Introduction}

Modeling beam dynamics in particle accelerators is key to both the design and operation of accelerator facilities. Accordingly, a number of beam tracking simulation codes have been developed over the years -- including Bmad, Elegant, Impact, ImpactX, MAD-X, Ocelot, SixTrack, Xsuite, just to name a few -- and are routinely used at various facilities. These codes track individual beam particles traveling through a series of accelerator lattice elements (quadrupoles, steering magnets, RF cavities, etc.) and many of these codes also incorporate the effect of 3D space charge. Although incorporating space charge significantly increases computational costs, it is crucial for capturing phenomena such as intensity-dependent beam losses~\cite{ElfredPRAB2021}, space-charge-driven emittance growth~\cite{YasuiPRAB2020}, as well as certain types of collective beam instabilities~\cite{QiangPRAB2018,ZolkinPRAB2018}.

In recent years, several \emph{differentiable} beam tracking codes have been developed~\cite{KaiserPRAB2024,QiangPRAB2023,GonzalezIPAC2023,KuklevConf2025,Wan2025}. Beyond computing output quantities (e.g., final beam size) from given inputs (e.g., quadrupole strengths, initial beam parameters, etc.), \emph{differentiable} tracking codes can also calculate the derivatives of those outputs with respect to the inputs. As such, differentiable tracking codes have enabled a number of new workflows, including gradient-based accelerator lattice optimization~\cite{KaiserPRAB2024}, gradient-based identification of unmeasured lattice misalignements~\cite{KaiserPRAB2024}, and more efficient variants of Bayesian optimization~\cite{KaiserPRAB2024,BoltzSciRep2025}. These differentiable codes also enable training neural networks \emph{while they are embedded} within a broader simulation pipeline, as is the case in generative phase space reconstruction~\cite{RousselPRL2023,RousselPRAB2024,HooverPRR2024}. Furthermore, they can also be used to study the stability of dynamical systems~\cite{Qiang2025a,Qiang2025b}. Finally, differentiable codes could also be leveraged for more efficient \emph{Simulation-Based Inference} (SBI)~\cite{CranmerPNAS2020}.

The abovementioned differentiable tracking codes make use of \emph{automatic  differentiation} -- a technique that enables calculation of the derivatives to machine precision (as opposed to approximate numerical methods for derivatives, such finite difference) without requiring developers to manually implement derivative calculations in the source code. Automatic differentiation itself can be performed using either the \emph{forward-mode} or \emph{reverse-mode} technique~\cite{Griewank2008,DorigoReviews2023}. In \emph{forward-mode} differentiation, each variable $a$ is paired with its gradient with respect to the inputs ($\partial a/\partial x_i$ for each input $x_i$), and each arithmetic operation on this variable (e.g., addition, multiplication) is overloaded to simultaneously carry out the corresponding chain-rule operation for its gradients. As a result, the required computational resources (both in time-to-solution and memory usage) scale linearly with the number of input variables.

By contrast, in \emph{reverse-mode} differentiation, the computation of derivatives is carried out in two distinct ``passes''. In the ``forward pass'', the simulation code is run using ordinary arithmetic operations to compute the output $f$, while recording data required for later gradient calculations. Then in the ``backward pass'', each operation is revisited in reverse order -- from output $f$ to the inputs $x_i$ -- going through each intermediate variable $a$ and computing $\partial f/\partial a$ until reaching $\partial f/\partial x_i$~\cite{Griewank2008,DorigoReviews2023} (see App.~\ref{app:illustrate} for an illustration based on a simple example).
As a result, the scaling of time-to-solution is largely independent of the number of input variables, making the reverse mode faster than the forward mode when handling a large number of inputs. This is especially valuable when training neural networks while they are embedded in a simulation \cite{RousselPRL2023,RousselPRAB2024,HooverPRR2024}, a task that requires finding gradients with respect to thousands or even millions of inputs (the weights of the neural network). However, one potential issue of reverse-mode differentiation is its memory usage. The memory usage indeed increases steadily during the forward pass, because of the need to continuously record data that will be used during the backward pass. As a result, there is a possibility that the simulation may run out of memory before the forward pass is completed.

The aim of this paper is thus to understand the scaling of memory usage in reverse-mode differentiation, in the particular case of beam tracking simulation with space charge. We study this scaling with respect to the number of macroparticles in the simulation, and the number of grid cells used for the space charge calculation. In order to carry out this study, we added support for space charge in the code Cheetah, an open-source beam tracking code that leverages PyTorch's built-in support for reverse-mode differentiation. (For more details on the Cheetah code itself, see~\cite{KaiserPRAB2024}.) To our knowledge, this is the first open-source beam tracking code which supports space charge with reverse-mode differentiation. The space-charge algorithm implemented in Cheetah uses standard Particle-In-Cell methods and is detailed in Sec.~\ref{sec:breakdown} ; benchmarks that confirm its accuracy are summarized in App.~\ref{app:benchmarks}.

While we measure the memory scaling specifically for the code Cheetah, the big $O$ scaling (e.g., whether the memory increases linearly or quadratically with the number of macroparticles) should not be specific to Cheetah and should apply to any beam tracking code with reverse-mode differentiation. Ultimately, these results can serve to estimate the memory requirements for a given differentiable beam tracking simulation, and whether this simulation is feasible on a given piece of computing hardware.

\section{Structure of beam tracking codes with space charge, and memory implications}
\label{sec:structure}

To gain a clearer understanding of how memory usage scales, it is helpful to first review the basic operating principles of beam tracking codes. These codes simulate the motion of $N_{\mathrm{part.}}$ macroparticles of a charged particle beam -- each often representing a large number of physical particles -- through a sequence of $N_{\mathrm{elements}}$ lattice elements. Importantly, rather than parameterizing trajectories using time $t$, these simulations typically employ the path length $s$ along the accelerator lattice as the independent variable. As the macroparticles advance in $s$, the algorithm updates their coordinates, which include the transverse positions $x$ and $y$, the relative arrival time $t$, the transverse momenta $p_x$, $p_y$, and the energy deviation $\delta E = E - E_0$ of each macroparticle -- where $E_0$ is the energy along the designed \emph{reference} trajectory for the accelerator lattice. (Depending on the specifics of each code, normalized versions of these coordinates may be used in practice.)

In cases where space-charge is neglected, each macroparticle is evolved \emph{independently} through the lattice elements. This uses a \emph{map} that gives the macroparticle coordinates at the exit of the element (at $s_{\mathrm{exit}}$) as function of those at the entrance (at $s_{\mathrm{entrance}}$). For simple enough representations of the lattice elements, the map often consists in a known, closed-form analytical expression. 

On the other hand, when space-charge is taken into account, there is no such analytical expression. A common approach is to employ second-order Strang splitting (with higher-order variants also in use). In this scheme, each lattice element is divided into $N_{\mathrm{slices}}$ segments of length $\Delta s$. For each slice, particles are first advanced without space-charge over $\Delta s / 2$ (independently, using maps), then their canonical momenta are updated to account for the space-charge interaction over $\Delta s$ at fixed $x, y, t$, and finally they are advanced again without space-charge for the remaining $\Delta s / 2$. The effect of space-charge over $\Delta s$ is commonly referred to as a ``space charge kick''. This is typically calculated using the Particle-In-Cell (PIC) method: macroparticles deposit their charge density onto a spatial grid, the fields are then computed on that grid by solving Poisson's equation, and the resulting Lorentz forces are interpolated back to the macroparticle positions to update their momenta. Further details are provided in Sec.~\ref{sec:breakdown}.

\begin{figure*}
    \centering
    \includegraphics[width=\linewidth]{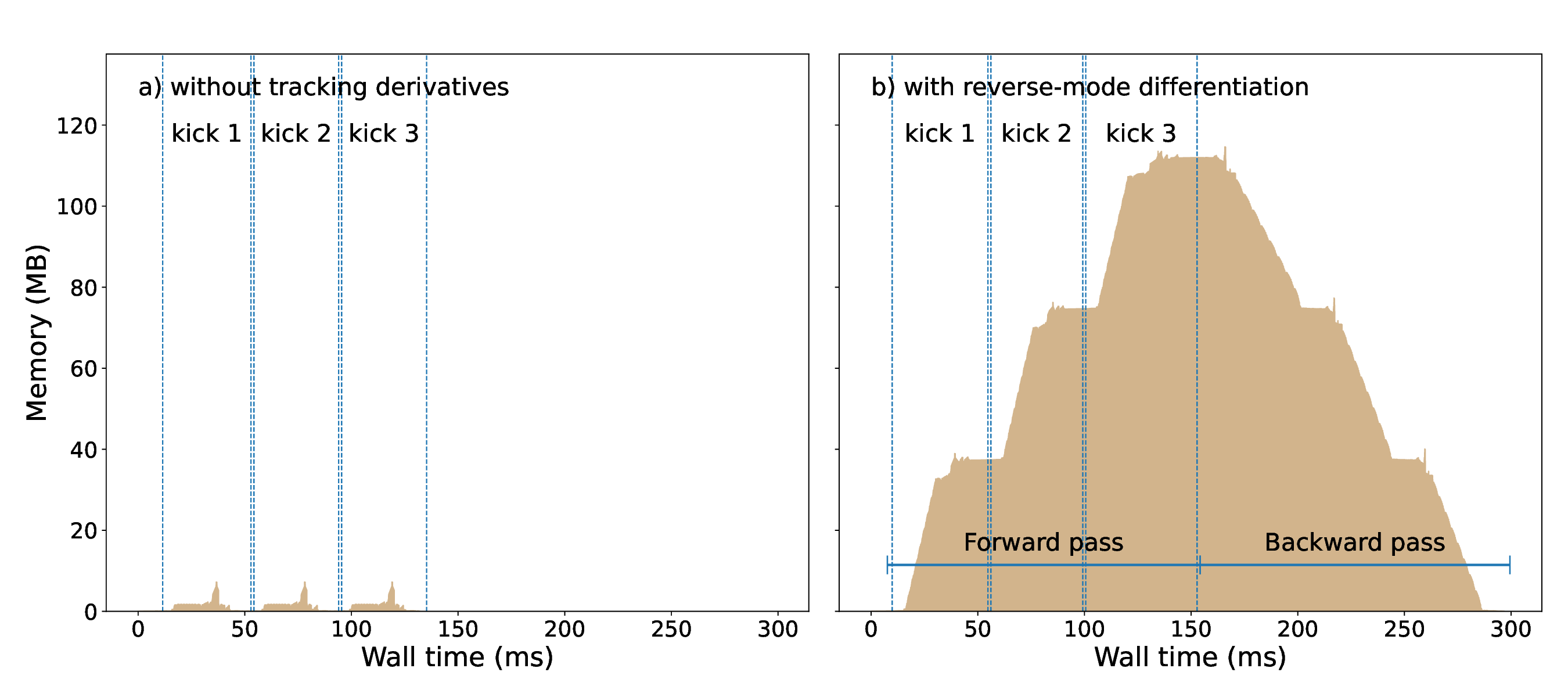}
    \caption{Timeline of memory usage for a Cheetah simulation (a) without tracking derivatives (left panel) and (b) with reverse-mode differentiation (right panel).}
    \label{fig:timeline}
\end{figure*}

In order to illustrate the typical behavior of memory usage in these simulations, we consider an electron bunch traveling through a drift (i.e., free space propagation) and expanding under its own space-charge. This setup is similar to the test labeled ``Free Expansion of a Cold Uniform Density Bunch'' in~\cite{MitchellICFA2024}, and the initial electron bunch is a uniform ellipsoid defined by a transverse radius $R_{i}$ and length $L_i$ (such that particles uniformly fill the space defined by $(x^2 + y^2)/R_{i}^2 + z^2/L_i^2 < 1$). The drift is subdivided into $N_{\mathrm{slices}}=3$ slices to take into account the effect of space-charge; the physical and numerical parameters of the simulation are summarized in Table~\ref{tab:parameters}. The simulation is run with the code Cheetah~\cite{KaiserPRAB2024} on an 80GB NVIDIA A100 GPU on the NERSC Perlmutter system. Fig.~\ref{fig:timeline} shows a timeline of memory usage, when the code is run without tracking derivatives (left panel in Fig.~\ref{fig:timeline}), or with derivatives (right panel in Fig.~\ref{fig:timeline}) -- in which case we use reverse-mode differentiation to compute the derivatives of the final RMS beam size $\sigma_{x, f}$ with respect to the initial beam radius $\partial \sigma_{x, f}/\partial R_i$, the initial beam length $\partial \sigma_{x, f}/\partial L_i$, the beam charge $\partial \sigma_{x, f}/\partial Q$, and the beam energy $\partial  \sigma_{x, f}/\partial \mathcal{E}$. All four derivatives are calculated in a single backward pass. The details of how memory usage was measured are summarized in App.~\ref{app:memory}.

\begin{table}
\begin{center}
  \begin{tabular}{cc}
    \hline
    Parameter & Value\\
    \hline
    Beam charge $Q$ & 10 nC \\
    Beam energy $\mathcal{E}$ & 250 MeV \\
    Beam initial radius $R_i$ & 1 mm \\
    Beam initial length $L_i$ & 1 mm \\
    Length of the drift $L$ & 5.5 m \\
    \hline
    Number of macroparticles $N_{\mathrm{part.}}$ & 1,000 \\
    Number of grid cells $N_{\mathrm{cells}}$ & $32^3$ \\
    Numer of slices $N_{\mathrm{slices}}$ & 3\\
    \bottomrule
  \end{tabular}
\end{center}  
  \caption{Physical and numerical parameters of the simulation}
  \label{tab:parameters}
\end{table}

Figure~\ref{fig:timeline} illustrates the difference in memory behavior with and without derivative tracking. In the non-differentiable case (left panel), memory is allocated and freed during each space charge kick, with no net growth over time. Closer inspection -- not shown here --indicates that, in this particular simulation, most of this transient allocation is related to computing temporary grid-based quantities used in the space-charge calculation.

In contrast, when reverse-mode differentiation is enabled (right panel), memory usage increases steadily during the forward pass and then decreases during the backward pass. As noted in the introduction and illustrated in App.~\ref{app:illustrate}, this behavior is expected: the code continuously records necessary data during the forward pass, which is then used to compute derivatives during the backward pass.

The figure further shows that, in the forward pass, memory usage increases by the same increment during each space charge kick (i.e., within each slice). This is again expected, as each space charge kick executes the same sequence of operations, leading to the same amount of data being stored for the backward pass. Consequently, the peak memory usage (i.e., the memory usage at the end of the forward pass) should scale proportionally to $N_{\mathrm{kicks}}$, the number of space charge kicks applied throughout a simulation. (When the simulation contains several lattice elements, $N_{\mathrm{kicks}}$ is sum of the number of slices for each element.) We expect that this scaling behavior is not unique to Cheetah and that it should apply generally to any beam-tracking code implementing reverse-mode differentiation.

Although the scaling with $N_{\mathrm{kicks}}$ is straightforward, the scaling with the number of macroparticles $N_{\mathrm{part.}}$ and grid cells $N_{\mathrm{cells}}$ is less apparent. Yet understanding this scaling is essential for estimating whether a simulation will exceed available memory by the end of the forward pass. This topic is examined in detail in Sec.~\ref{sec:scaling}.

Before turning to that analysis, it is useful to make a few additional observations in Fig.~\ref{fig:timeline}. First of all, the evaluation of space charge kicks accounts for most of the computational cost, both for the simulation without derivatives (left panel) and for the forward pass of the differentiable simulation (right panel). The small gaps between the dashed lines for consecutive kicks correspond to the application of linear maps to each particle (as required in the Strang splitting scheme) and take negligible time. These small gaps also do not incur any perceptible change in memory consumption, compared to the space charge kicks, and this is the reason why this paper focuses on the memory usage due to the space charge kick, rather than that due to the application of linear maps.
Secondly, the forward pass in the differentiable simulation (right panel) is only marginally slower than the non-differentiable run. This is indeed one of the advantages of reverse-mode differentiation, compared to forward-mode differentiation.

\section{Detailed memory scaling of the space charge kick with number of macroparticles and cells}
\label{sec:scaling}

\subsection{Overall memory scaling}

In order to study memory usage during the forward pass with respect to $N_{\mathrm{part.}}$ and $N_{\mathrm{cells}}$, we now run a \emph{single} space charge kick. We use the same physical parameters as in Table~\ref{tab:parameters}, but we systematically vary $N_{\mathrm{part.}}$ and $N_{\mathrm{cells}}$. The increase in memory usage over one single space charge kick is measured as described in App.~\ref{app:memory}.

\begin{figure}
    \centering
    \includegraphics[width=\linewidth]{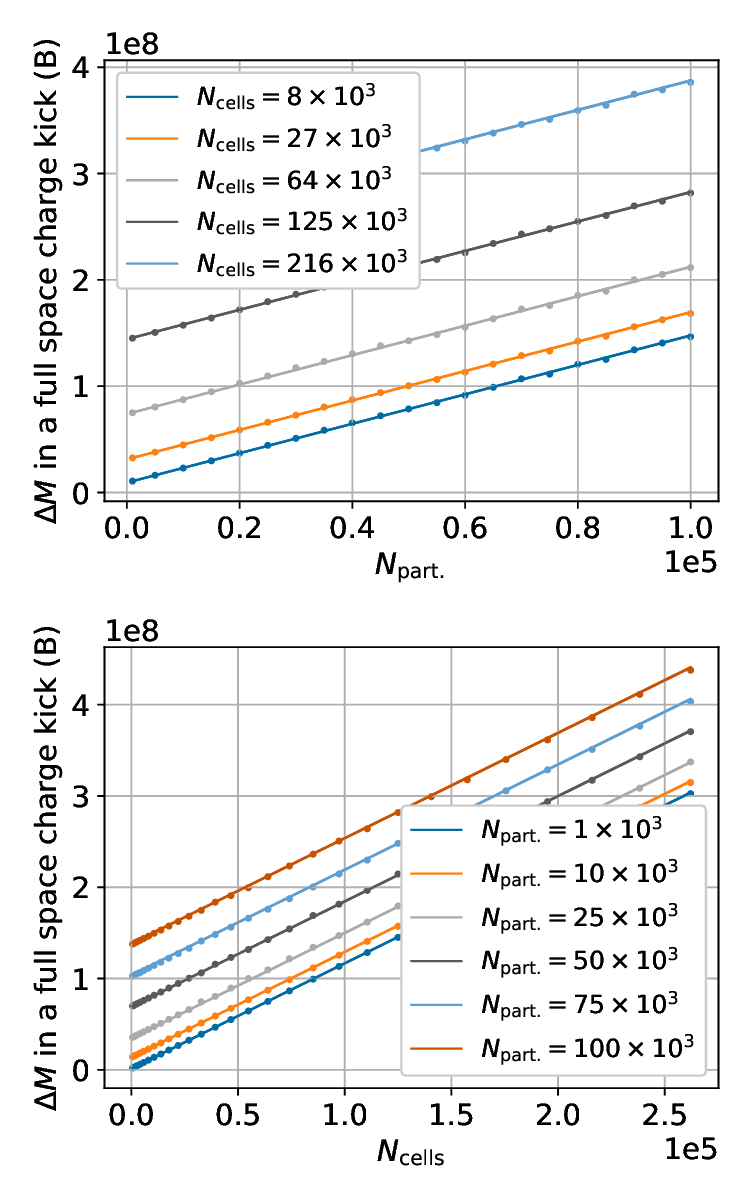}
    \caption{Total change in memory usage over one space charge kick (forward pass, with reverse-mode differentiation turned on), as a function of the number of macroparticles and cells. Eq.~(\ref{eq:linear}) is fitted to the data, and the corresponding coefficients are represented in Fig.~\ref{fig:coefficients}. The lines show the evaluation of this fitted formula for fixed number of cells (top) and fixed number of macroparticles (bottom).}
    \label{fig:total}
\end{figure}

\begin{figure}
    \centering
    \includegraphics[width=\linewidth]{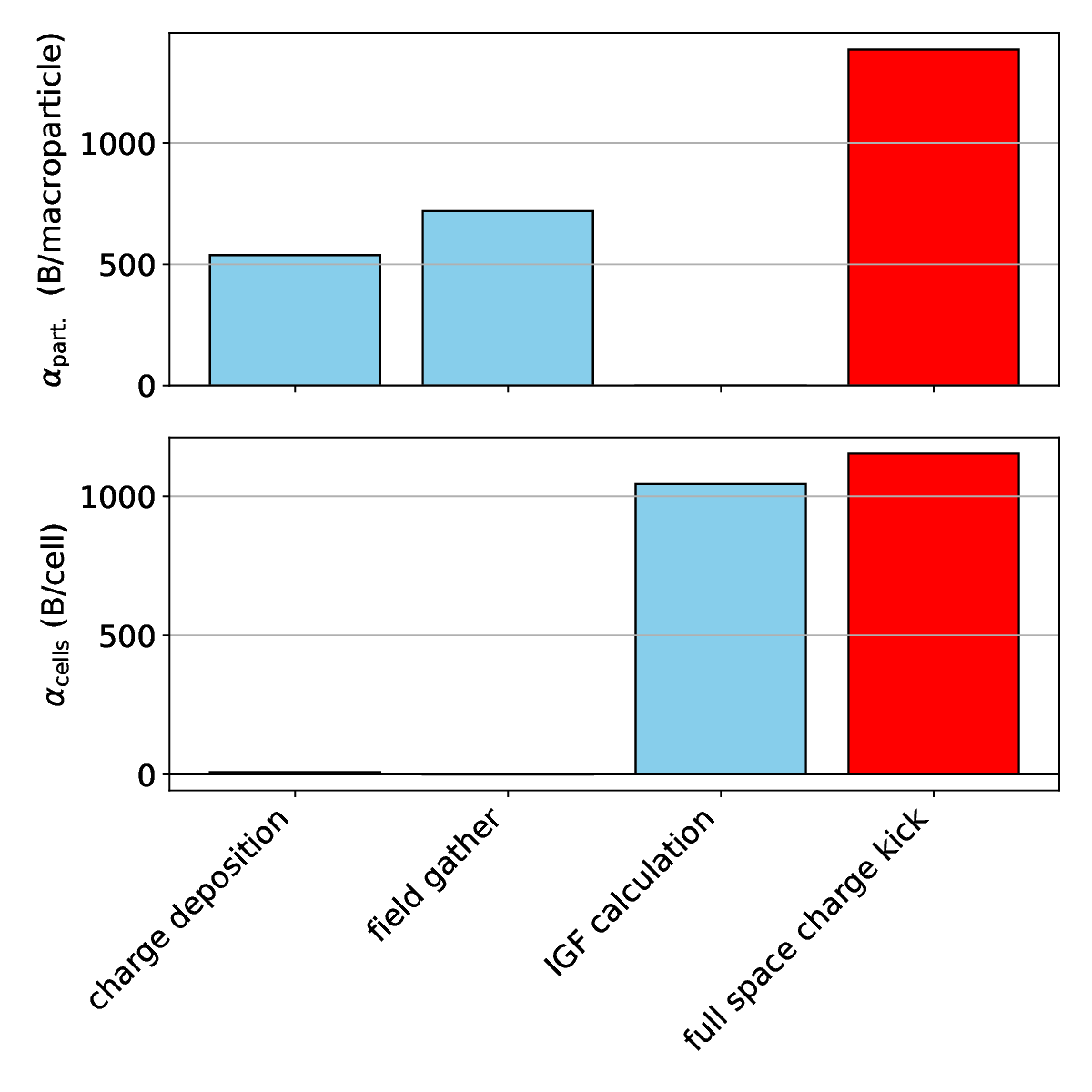}
    \caption{Coefficients for the increase in memory usage during the forward pass, as given by Eq.~(\ref{eq:linear}), when applied to a full space charge kick (red bars) or to \emph{selected} parts of the space charge kick (blue bars).}
    \label{fig:coefficients}
\end{figure}

The result of this scan over $N_{\mathrm{part.}}$ and $N_{\mathrm{cells}}$ is displayed in Fig.~\ref{fig:total}, where it is compared with a linear fit of the form 
\begin{equation}
\Delta M = \alpha_{\mathrm{part.}}N_{\mathrm{part.}}+\alpha_{\mathrm{cells}}N_{\mathrm{cells}}
\label{eq:linear}
\end{equation}
where $\Delta M$ is the increase in memory usage (in bytes) after a single space charge kick -- during the forward pass, with reverse-mode differentiation turned on. (See App.~\ref{app:fit} for more details on how this linear fit was performed.) The corresponding values of the coefficients $\alpha_{\mathrm{part.}}$ and $\alpha_{\mathrm{cells}}$ are represented in Fig.~\ref{fig:coefficients} (red bars).

As can be seen in Fig.~\ref{fig:total}, the data points align with this linear fit, and thus the memory usage is linear in both $N_{\mathrm{part.}}$ and $N_{\mathrm{cells}}$. Indeed if there was any significant nonlinear term such as $N_{\mathrm{cells}}^2$ or $N_{\mathrm{part.}}^2$, we would see a clear departure of the data points from the linear fits in Fig.~\ref{fig:total}. Similarly, if there was any significant nonlinearity of the form $N_{\mathrm{cells}}\times N_{\mathrm{part.}}$, the data points would align along lines that would have different \emph{slopes} for different values of $N_{\mathrm{cells}}$ on the top panel of Fig.~\ref{fig:total} and for different values of $N_{\mathrm{part.}}$ on the bottom panel of Fig.~\ref{fig:total}, which is not the case.

The fact that the memory scaling is linear is encouraging for the practical applicability of differentiable beam tracking code. Any nonlinear term (e.g., $N_{\mathrm{cells}}^2$, $N_{\mathrm{part.}}^2$, $N_{\mathrm{cells}}\times N_{\mathrm{part.}}$) could indeed have led to prohibitively large memory usage, given the typically large values of $N_{\mathrm{cells}}$ and $N_{\mathrm{part.}}$. For instance, it is not uncommon to use on the order of $10^5-10^6$ macroparticles, and similar numbers for the number of cells. Note that, one could have \emph{a priori} expected such nonlinear terms to exist. For instance, a space charge kick involves computing the potential $\phi$ from the charge density $\rho$, on a grid (see Sec.~\ref{sec:breakdown}). Since this involves solving an elliptic equation, in principle $\partial \phi_{ijk}/\partial \rho_{i'j'k'}$ is non-zero for any pair of cells $(ijk, i'j'k')$. One could have expected that this matrix of size $N_{\mathrm{cells}}\times N_{\mathrm{cells}}$ would need to be stored during the forward pass, so as to propagate gradients during the backward pass.

In order to better understand the observed linear scaling, in the rest of this section we break down the space charge kick into its sub-steps (Sec.~\ref{sec:breakdown}) and inspect the memory scaling of some key sub-steps (Sec.~\ref{sec:field_solver}, \ref{sec:charge_deposition}, \ref{sec:field_gather}).

\subsection{Breakdown of the algorithmic steps within one space charge kick}
\label{sec:breakdown}

As mentioned previously, the purpose of the space charge kick is to update the momentum of each macroparticle to capture the impact of space charge over a distance $\Delta s$. In Cheetah, the space charge kick consists of the following steps, which are similar to the algorithm used in other beam tracking codes:
\begin{itemize}
\item \textbf{$s$-to-$t$ transform:} The macroparticles are converted from a representation at fixed $s$ (where each macroparticle is represented with $x$, $y$, $t$, $p_x$, $p_y$, $\delta E$) to a representation at fixed time (where each macroparticle is represented with $x$, $y$, $z=-\beta ct$, $p_x$, $p_y$, $p_z=\sqrt{ (E_0 + \delta E)^2/c^2 - p_x^2 - p_y^2 - m^2c^2}$).
\item \textbf{Charge deposition:} A Cartesian grid that overlaps with the beam is created (with cell size $\Delta x, \Delta y, \Delta z$), and the charge density $\rho$ is computed on the grid using the Cloud-In-Cell method \cite{Birdsall}, i.e. for each grid point indexed by $ijk$,
\[ \rho_{ijk} = \frac{1}{\Delta x \Delta y \Delta z}\sum_{p=1}^{N_{part.}} q_p S(\boldsymbol{x}_{ijk}-\boldsymbol{x}_p) \]
where $q_p$ and $\boldsymbol{x}_p$ is the charge and position of macroparticle $p$, $\boldsymbol{x}_{ijk}$ is the position of grid point $ijk$, and $S$ is the product of first-order spline functions in $x$, $y$ and $z$ \cite{Birdsall}.
\item \textbf{Field solver:} For a relativistic beam propagating along the $z$ axis with a Lorentz factor $\gamma$, the Lorentz force $\boldsymbol{F} = q(\boldsymbol{E} + \boldsymbol{v}\times\boldsymbol{B})$ can be obtained with $\boldsymbol{F} = -q \boldsymbol{\nabla}\phi/\gamma^2$ where $\phi$ is found by solving a modified Poisson equation \cite{VayPoP2008} 
\begin{equation}
\partial_x^2\phi + \partial_y\phi^2 + \frac{1}{\gamma^2}\partial_z^2\phi = -\frac{\rho}{\epsilon_0}. \label{eq:modPoisson}
\end{equation}
Cheetah solves this Poisson equation by using the Integrated Green Function (IGF) method \cite{QiangPRAB2006} to obtain $\phi$ on the grid, and then computes $\boldsymbol{F}= -q\boldsymbol{\nabla}\phi/\gamma^2$ on the same grid points using centered finite differences.
\item \textbf{Field gather:} The Lorentz force $\boldsymbol{F}$ is interpolated from the grid points to the macroparticles using:
\[ \boldsymbol{F}(\boldsymbol{x}_p) = \sum_{ijk}S(\boldsymbol{x}_p - \boldsymbol{x}_{ijk})\boldsymbol{F}_{ijk}\]
\item \textbf{Momentum update:} The momentum of each macroparticle $\boldsymbol{p}=(p_x, p_y,p_z)$ is updated using
\[ \boldsymbol{p} \leftarrow \boldsymbol{p} + \boldsymbol{F}\Delta t\]
where $\boldsymbol{F}$ is the interpolated Lorentz force and $\Delta t = \Delta s/\beta_z c$ is the amount of time it takes the beam to travel the distance $\Delta s$.
\item \textbf{$t$-to-$s$ transform:} The energy deviation $\delta E$ is obtained with $\delta E = c\sqrt{p_x^2 + p_y^2 + p_z^2 + m^2c^2} - E_0$.
\end{itemize}
Note that the approach of using the modified Poisson equation Eq.~(\ref{eq:modPoisson}) is mathematically equivalent to solving the regular Poisson equation in the beam's rest frame, and transforming the Lorentz force back to the laboratory frame -- which is the approach used in some of the other beam tracking codes.

The $t$-to-$z$ transform, $z$-to-$t$ transform and momentum update correspond to a series of element-wise operations on macroparticle arrays of size $N_{\mathrm{part.}}$ (element-wise multiplications, additions, square root), and thus it seems natural that the memory increase incurred during these parts of the forward pass scales as $O(N_{\mathrm{part.}})$. (Again see App.~\ref{app:illustrate} for an example of why memory increases during the forward pass.) On the other hand, the scaling of the charge deposition, field solver and field gather is less intuitive and will be examined in more detail in the following subsections.

\subsection{Memory scaling of the field solver}
\label{sec:field_solver}

The field solver uses the Integrated Green Function (IGF) method \cite{QiangPRAB2006}, whereby $\phi$ is given by the convolution of $\rho$ with the IGF \cite{QiangPRAB2006}:
\begin{equation}
\phi_{ijk} = (\rho * G)_{ijk} = \sum_{i'j'k'}\rho_{i'j'k'}G(\boldsymbol{x}_{ijk} - \boldsymbol{x}_{i'j'k'})
\label{eq:convolution}
\end{equation}
The IGF $G(\boldsymbol{x})$ is the integral of the Green function for Eq.~(\ref{eq:modPoisson}), $\phi(x, y, z) = \gamma/4\pi \epsilon_0\sqrt{x^2 + y^2 + \gamma^2 z^2}$, over one cell. It can be evaluated by taking sums and differences of a somewhat complicated analytical function $g(x,y,z)$ \cite{qiang2024implementation}:
\begin{align} 
&G(\boldsymbol{x}) = \nonumber\\
&\quad g\left(x+\frac{\Delta x}{2}, y+\frac{\Delta y}{2}, \gamma z+\frac{\gamma\Delta z}{2}\right) - g\left(x-\frac{\Delta x}{2}, y+\frac{\Delta y}{2}, \gamma z+\frac{\gamma\Delta z}{2}\right) \nonumber\\
&- g\left(x+\frac{\Delta x}{2}, y-\frac{\Delta y}{2}, \gamma z+\frac{\gamma\Delta z}{2}\right) - g\left(x+\frac{\Delta x}{2}, y+\frac{\Delta y}{2}, \gamma z-\frac{\gamma\Delta z}{2}\right) \nonumber\\
&+g\left(x+\frac{\Delta x}{2}, y-\frac{\Delta y}{2}, \gamma z-\frac{\gamma\Delta z}{2}\right) +g\left(x-\frac{\Delta x}{2}, y+\frac{\Delta y}{2}, \gamma z-\frac{\gamma\Delta z}{2}\right) \nonumber\\
&+g\left(x-\frac{\Delta x}{2}, y-\frac{\Delta y}{2}, \gamma z+\frac{\gamma\Delta z}{2}\right) -g\left(x-\frac{\Delta x}{2}, y-\frac{\Delta y}{2}, \gamma z-\frac{\gamma\Delta z}{2}\right)
\label{eq:igf_G}
\end{align}
with
\begin{align}
&g(x,y,z) = \frac{1}{4\pi \epsilon_0}\left[ yz\,\mathrm{arcsinh}\left(\frac{x}{\sqrt{y^2 +z^2}}\right) + xz \,\mathrm{arcsinh}\left(\frac{y}{\sqrt{x^2 +z^2}}\right)\right. \nonumber \\
&\qquad+xy \,\mathrm{arcsinh}\left(\frac{z}{\sqrt{x^2 +y^2}}\right) - \frac{z^2}{2}\arctan\left(\frac{xy}{z\sqrt{x^2+y^2+z^2}}\right) \nonumber\\
&\left. \quad- \frac{y^2}{2}\arctan\left(\frac{xz}{y\sqrt{x^2+y^2+z^2}}\right) - \frac{x^2}{2}\arctan\left(\frac{yz}{x\sqrt{x^2+y^2+z^2}}\right)\right]
\label{eq:igf_g}
\end{align}

A direct implementation of the convolution in Eq.~(\ref{eq:convolution}) would involve $O(N_{\mathrm{cells}}^2)$ operations. Instead, Cheetah (like many other beam tracking codes) uses the Hockney method \cite{Hockney} to evaluate Eq.~(\ref{eq:convolution}). This method leverages the fact that convolutions can be evaluated efficiently using FFTs, and consists in the following steps:
\begin{itemize}
\item Compute the IGF $G(\boldsymbol{x})$ on the spatial grid using Eqs.~(\ref{eq:igf_G}) and (\ref{eq:igf_g})
\item Compute the Fourier transform of $G$ and $\rho$, using the FFT algorithm: $\hat{G} = FFT(G)$, $\hat{\rho} = FFT(\rho)$.
\item Perform the element-wise multiplication of $\hat{\rho}$ and $\hat{G}$ in Fourier space: $\hat{\phi} = \hat{\rho}\hat{G}$
\item Obtain $\phi$ by computing the inverse Fourier transform of $\hat{\phi}$: $\phi = FFT^{-1}(\hat{\phi})$
\end{itemize}

\begin{figure}
    \centering
    \includegraphics[width=\linewidth]{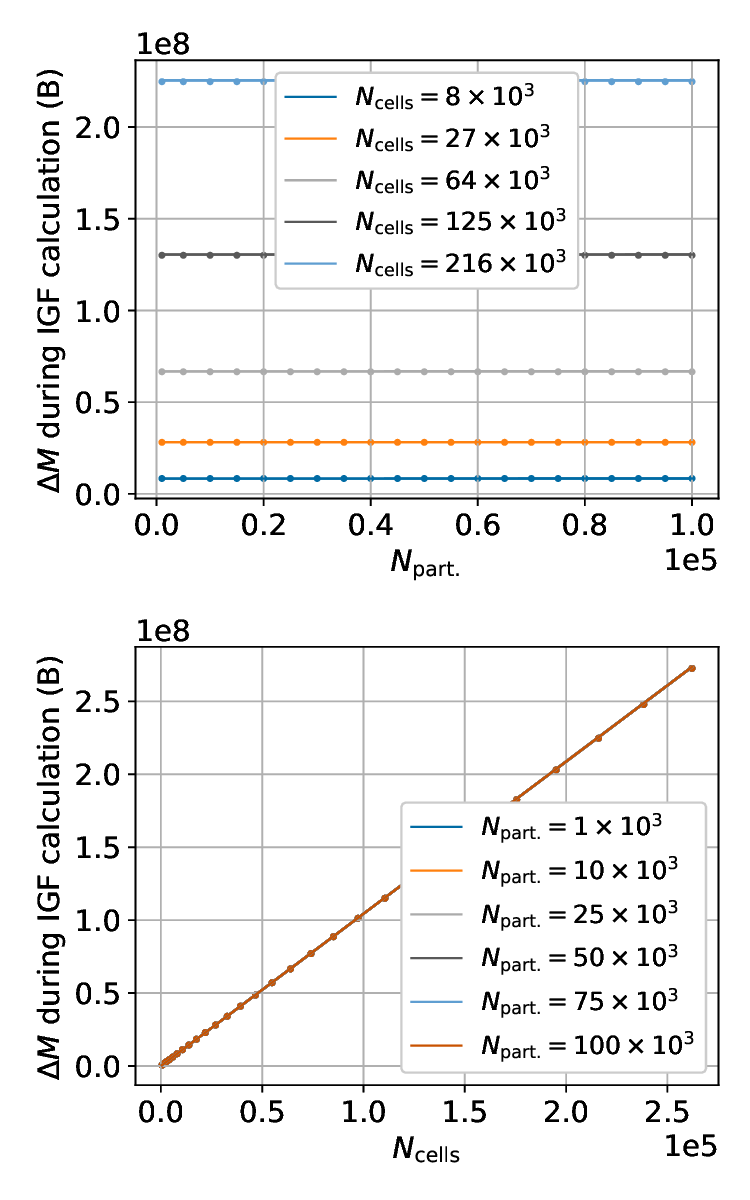}
    \caption{Increase in memory usage $\Delta M$ incurred during the calculation of the Integrated Green Function (IGF), as part of one space charge kick (forward pass, with reverse-mode differentiation turned on), as a function of the number of macroparticles and cells. Eq.~(\ref{eq:linear}) is fitted to the data, and the corresponding coefficients are represented in Fig.~\ref{fig:coefficients}. The lines show the evaluation of this fitted formula for fixed number of cells (top) and fixed number of macroparticles (bottom).}
    \label{fig:green_function_calculation}
\end{figure}

The first step in the above (computing the IGF $G(\boldsymbol{x})$ on the spatial grid) consists in a series of \emph{element-wise} operations (given by Eq.~(\ref{eq:igf_G}) and Eq.~(\ref{eq:igf_g})) on arrays of size $N_{cells}$. It is thus expected that the size of the data recorded during the forward pass scales proportionally to $N_{cells}$ -- and that it does not depend on $N_{\textrm{part.}}$, since the macroparticles are not involved in these operations. This is indeed confirmed in Fig.~\ref{fig:green_function_calculation}. Note from Fig.~\ref{fig:coefficients} that the coefficient $\alpha_{\text{cells}}$ for the IGF calculation (rightmost blue bar in the bottom panel) accounts for most of the memory usage per cell in the full space charge kick (red bar in the bottom panel). This can be explained by the large amount of individual operations involved in evaluations of Eq.~(\ref{eq:igf_G}) and Eq.~(\ref{eq:igf_g}), and the fact that, during the forward pass, some data  needs to be recorded for each of these individual operations.

Similarly, the third step in the Hockney method (multiplication of $\hat{\rho}$ and $\hat{G}$ in Fourier space) is also an element-wise operation on arrays of size $O(N_{\mathrm{cells}})$, and thus the amount of data recorded during the forward pass is again expected to scale proportionally to $N_{\mathrm{cells}}$. On the other hand, the second and fourth steps involve FFTs which are \emph{not} element-wise operations.

When measuring the change in memory usage during these FFTs and inverse FFT (using the tools described in App.~\ref{app:memory}), we found that \emph{no additional data} is recorded during these parts of the forward pass -- in other words $\Delta M = 0$ irrespective of $N_{\mathrm{cells}}$. This implies that no additional data is needed in order to propagate the gradients through an FFT, during the backward pass. Indeed a Fourier transform can be written as:
\[ \hat{F}_{\ell'} = \sum_{\ell=1}^N e^{-2i\pi \frac{\ell \ell'}{N}} F_{\ell}\]
and applying the chain rule to propagate the gradients during the backward pass yields:
\[
\frac{\partial f}{\partial F_\ell} = \sum_{\ell'=1}^N \frac{\partial f}{\partial \hat{F}_{\ell'}}\frac{\partial \hat{F}_{\ell'}}{\partial F_{\ell}} = \sum_{\ell'} \frac{\partial f}{\partial \hat{F}_{\ell'}} e^{-2i\pi \frac{\ell \ell'}{N}} 
\]
Hence the operation that propagates the gradients during the backward pass is itself a Fourier transform. (This can be confirmed for instance by inspecting the code of open-source packages that implement reverse-mode differentiation for FFTs \cite{pytorchfft}.) This operation does not require additional input data beyond the gradients to be propagated $\partial f/\partial \hat{F}_{\ell'}$ (compare this with some of the operations in App.~\ref{app:illustrate} that \emph{do} require additional input data to propagate gradients in the backward pass), and thus this explains why no additional data is recorded when performing FFTs during the forward pass.

Overall, this section showed that the field solver is a combination of element-wise operation (which cause memory usage to increase proportionally to $N_{\mathrm{cells}}$ during the forward pass) and of Fourier transforms (which cause no increase in memory usage during the forward pass). Taken together, these observations indicate that the increase in memory usage for the field solver, during the forward pass, scales like $O(N_{\mathrm{cells}})$.

It is worth noting that, beyond FFT-based solvers, real-space iterative solvers (e.g. multigrid solvers, GMRES, as well as multigrid-preconditioned Krylov solvers) are also commonly used in beam tracking codes -- due to their capacity to handle beam pipes with complex geometries. While the FFT operation does not incur \emph{any} memory increase during the forward pass, iterative solvers likely do not possess a similar property. Consequently, their memory usage scaling is expected to deviate from the linear $O(N_{\mathrm{cells}})$ behavior observed for FFT-based solvers. However, a comprehensive analysis of the memory scaling of iterative solvers falls outside the scope of this work.

\subsection{Memory scaling of the charge deposition}
\label{sec:charge_deposition}

\begin{figure}
    \centering
    \includegraphics[width=\linewidth]{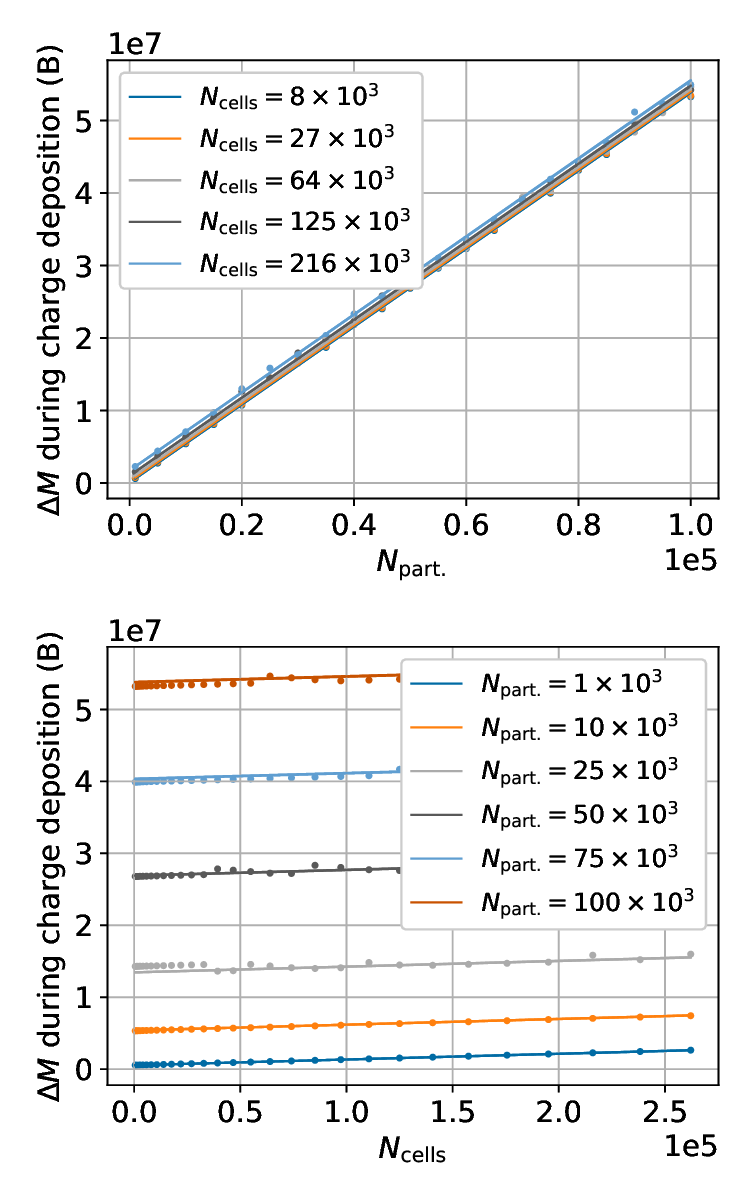}
    \caption{Increase in memory usage $\Delta M$ incurred during charge deposition, as part of one space charge kick (forward pass, with reverse-mode differentiation turned on), as a function of the number of macroparticles and cells. Eq.~(\ref{eq:linear}) is fitted to the data, and the corresponding coefficients are represented in Fig.~\ref{fig:coefficients}. The lines show the evaluation of this fitted formula for fixed number of cells (top) and fixed number of macroparticles (bottom).}
    \label{fig:charge_deposition}
\end{figure}

Fig.~\ref{fig:charge_deposition} shows the increase in memory usage measured during charge deposition (using the profiling tools described in App.~\ref{app:memory}) as part of the forward pass. As can be observed, the memory usage depends strongly on the number of macroparticles ($\alpha_{\mathrm{part.}} \simeq 550\,\mathrm{B}$ per macroparticle) and increases only mildly as a function of the number of cells ($\alpha_{\mathrm{cells}} \simeq 8\,\mathrm{B}$ per cell). We observe no evidence for a dependency of the form $N_{\mathrm{part.}}\times N_{\mathrm{cells}}$, despite the fact that the formula
\[ \rho_{ijk} = \frac{1}{\Delta x \Delta y \Delta z}\sum_{p=1}^{N_{part.}} q_p S(\boldsymbol{x}_{ijk}-\boldsymbol{x}_p) \]
involves both cells and particles. The reason is that this is a sparse operation, whereby $S(\boldsymbol{x}_{ijk}-\boldsymbol{x}_p)$ is a compact function which evaluates to zero for most grid points. When using Cloud-In-Cell functions for $S$ (i.e., first-order splines), each macroparticle indeed only contributes to its two nearest grid points along each dimension (i.e., it contributes to 8 grid points in 3D). In practice, in Cheetah, this is implemented with the following steps:
\begin{itemize}
\item Compute the weighting coefficients
\[ S(\boldsymbol{x}_{ijk}- \boldsymbol{x}_{p}) \equiv \left( 1 - \frac{|x_i - x_p|}{\Delta x} \right)\times\left( 1 - \frac{|y_j - y_p|}{\Delta y} \right)\times\left( 1 - \frac{|z_k - z_p|}{\Delta z} \right)\]
for the 8 grid points surrounding each macroparticle. This results in a set of $8\times N_{\mathrm{part.}}$ coefficients.
\item Add these contributions at the proper indices in the array $\rho$, using the PyTorch function \texttt{index\_put\_}. 
\end{itemize}
Since the above steps generally involve $O(N_{\mathrm{part.}})$ operations, it seems natural that the data recorded during the forward pass scales like $O(N_{\mathrm{part.}})$. Further investigation -- not shown here -- revealed that the weak dependency with $N_{\mathrm{cells}}$ comes from data recorded during auxiliary operations on the array $\rho$ (allocation, normalization by $1/\Delta x\Delta y\Delta z$).

\subsection{Memory scaling of the field gather}
\label{sec:field_gather}

The field gather operation is structurally similar to the charge deposition and consists in the following steps:
\begin{itemize}
\item Compute the weighting coefficients $S(\boldsymbol{x}_{ijk}- \boldsymbol{x}_{p})$ for the 8 grid points surrounding each macroparticle.
\item Sum the contributions to $\boldsymbol{F}(\boldsymbol{x}_p)$ from the 8 nearest grid point for each macroparticle, using the PyTorch function \texttt{scatter\_add}.
\end{itemize}

\begin{figure}
    \centering
    \includegraphics[width=\linewidth]{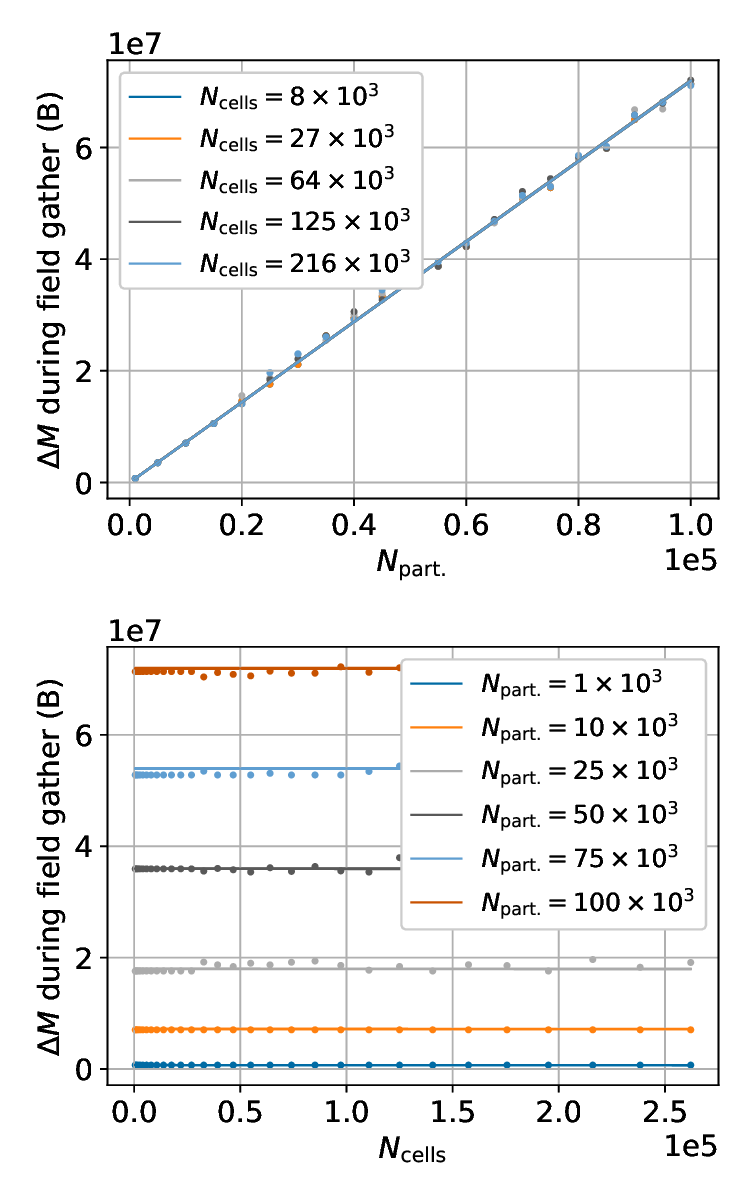}
    \caption{Increase in memory usage $\Delta M$ incurred during field gather, as part of one space charge kick (forward pass, with reverse-mode differentiation turned on), as a function of the number of macroparticles and cells. Eq.~(\ref{eq:linear}) is fitted to the data, and the corresponding coefficients are represented in Fig.~\ref{fig:coefficients}. The lines show the evaluation of this fitted formula for fixed number of cells (top) and fixed number of macroparticles (bottom).}
    \label{fig:field_gather}
\end{figure}

The increase in memory usage during field gather is shown in Fig.~\ref{fig:field_gather}. Again, we observe a strong dependency with $N_{\mathrm{part.}}$ (and in this case no dependency on $N_{\mathrm{cells}}$). As can be seen in Fig.~\ref{fig:coefficients}, charge deposition and field gather account for most of the memory consumption per macroparticle, during the forward pass of a full space charge kick.

\section{Conclusion}

In this paper, we studied the scaling of the increase in memory usage, during the forward pass of a beam tracking simulation using space charge with reverse-mode differentiation. Overall, we found that the memory scaling is linear in the number of macroparticles $N_{\mathrm{part.}}$ and cells $N_{\mathrm{cells}}$, and that it increases with the number of space charge kicks applied to the beam throughout the simulation $N_{\mathrm{kicks}}$:
\begin{equation} 
\Delta M \simeq (\alpha_{\mathrm{part.}}N_{\mathrm{part.}} + \alpha_{\mathrm{cells}}N_{\mathrm{cells}})\times N_{\mathrm{kicks}}
\end{equation}

The fact that the memory usage grows linearly in $N_{\mathrm{part.}}$ and $N_{\mathrm{cells}}$ (and not, e.g., quadratically) suggests that the memory cost of space charge with reverse-mode differentiation is not prohibitively expensive -- provided that the number of discrete space charge kicks $N_{\mathrm{kicks}}$ is not too large. For applications that require a large number of space charge kicks (such as beam tracking in a ring over thousands of turns or more), the issues of growing memory usage can potentially be mitigated by using \emph{gradient checkpointing} \cite{pytorch2024checkpoint}, a technique that reduces memory consumption at the expense of additional computation -- by storing only a limited subset of data during the forward pass (in special checkpoints) and recomputing portions of the forward pass from these checkpoints during the backward pass.

The PIC methods employed here (namely charge deposition and field gather with compact shape factors, and FFT-based convolution for the field solve) are standard components shared by many beam tracking codes. As such, we expect the finding of linear memory scaling with simulation resolution to generalize beyond Cheetah to any code implementing FFT-based space charge with reverse-mode differentiation. This expectation extends naturally to other collective effects, such as wakefields or coherent synchrotron radiation, which are typically handled with analogous numerical techniques. In this sense, the linear scaling law reported here should be understood as a general property of reverse-mode differentiation applied to FFT-based collective effect solvers, rather than a result specific to Cheetah.

Nevertheless, the numerical values of the memory coefficients are, of course, implementation-dependent. In the specific case of Cheetah, we find that the memory overhead per space charge kick (see Fig.~\ref{fig:coefficients}) amounts to $\alpha_{\mathrm{part.}} = 1.4\,\mathrm{kB}$ per macroparticle — dominated by data recorded during charge deposition and field gather -- and $\alpha_{\mathrm{cells}} = 1.2\,\mathrm{kB}$ per cell, dominated by the evaluation of the IGF. While these specific values apply to the current version of Cheetah and will differ across codes and implementations, they serve to illustrate how the linear scaling law can be used in practice to predict memory consumption and assess feasibility before running a simulation. Furthermore, regardless of the specific implementation, this work suggests that future memory optimization efforts, in Cheetah or in other codes, should prioritize the evaluation of the IGF analytical expression, as well as the charge deposition and field gather operations, since these consistently appear to dominate the memory cost of reverse-mode differentiation.

\begin{acks}

This work was supported
by the Laboratory Directed Research and Development Program of
Lawrence Berkeley National Laboratory, as well as by the U.S. Department of Energy, Office
of Science, Office of High Energy Physics, General Accelerator
R\&D (GARD), under Contract No. DE-AC02-05CH11231. 
This research used resources of the National Energy Research Scientific Computing Center, a DOE Office of Science User Facility supported by the Office of Science of the U.S. Department of Energy under Contract No. DE-AC02-05CH11231.
This work was partially funded in the context of the 205 R\&D program of the European XFEL.
\end{acks}

\section{Data availability statement}

The scripts and instructions needed to reproduce these results are available on Zenodo.org at https://doi.org/10.5281/zenodo.17822325.

This research used the following software versions:

\begin{itemize}
\item CUDA 12.6
\item CPython 3.13.5 
\item PyTorch 2.7.1
\item A modified version of Cheetah v0.7.4, which was instrumented for memory profiling (see the Zenodo archive)
\end{itemize}

\appendix

\section{Illustration of the memory requirements of reverse-mode differentiation, in a simple example}
\label{app:illustrate}

In order to illustrate why memory usage increases during the forward pass of reverse-mode differentiation, we consider a simple example:
\[ f(\boldsymbol{x}) = || \log(M\boldsymbol{x}) ||^2 \]
where $\boldsymbol{x}$ is a vector of size $N$, $M$ is a matrix of size $N\times N$, and where we wish to compute the derivative of $f$ with respect to $\boldsymbol{x}$, $\partial f/\partial x_i$ for $i\in[1,N]$.

In this case, the forward pass proceeds in three steps:
\begin{itemize}
\item Step 1: compute $\boldsymbol{a} = M\boldsymbol{x}$ (matrix multiplication), and keep $M$ in memory ($N \times N$ scalars)
\item Step 2: compute $\boldsymbol{b} = \log( \boldsymbol{a} )$ (element-wise operation), and keep the $1/a_i$ in memory ($N$ scalars)
\item Step 3: compute $f = ||\boldsymbol{b}||^2 \equiv \sum_{i=1}^N b_i^2$, and keep $\boldsymbol{b}$ in memory ($N$ scalars)
\end{itemize}

The reason why different variables needed to be kept in memory becomes apparent when considering the backward pass. In the backward pass, the previous steps are revisited in reverse order, to calculate the derivatives of $f$ with respect to $\boldsymbol{b}$, $\boldsymbol{a}$ and finally $\boldsymbol{x}$ using the chain rule:
\begin{itemize}
\item From step 3: $\frac{\partial f}{\partial b_i} = 2 b_i$, which is known because $\boldsymbol{b}$ was kept in memory. After computing $\frac{\partial f}{\partial b_i}$, the memory that held $\boldsymbol{b}$ can be deallocated.
\item From the chain rule applied to step 2: $\frac{\partial f}{\partial a_i} = \frac{\partial f}{\partial b_i}\frac{\partial b_i}{\partial a_i}$. $\frac{\partial f}{\partial b_i}$ is known from the previous step of the backward pass, and by property of the $\log$, $\frac{\partial b_i}{\partial a_i} = \frac{1}{a_i}$, which is known since the $1/a_i$ were kept in memory. After computing $\frac{\partial f}{\partial a_i}$, the memory that held the $1/a_i$ can be deallocated.
\item From the chain rule applied to step 1: $\frac{\partial f}{\partial x_i} = \sum_{j=1}^N\frac{\partial f}{\partial a_j}\frac{\partial a_j}{\partial x_i}$. $\frac{\partial f}{\partial a_j}$ is known from the previous step of the backward pass, and by property of the matrix multiplication, $\frac{\partial a_j}{\partial x_i} = M_{ji}$, which is known because the matrix $M$ was kept in memory. After computing $\frac{\partial f}{\partial x_i}$, the memory that held $M$ can be deallocated.
\end{itemize}

Of course the above is a very simple example, and a full simulation consists of a much longer chain of operations. But conceptually each of them can be thought of as going from a set of intermediate variables $\{ a_i \}$ to a new set of intermediate variables $\{ a_j' \}$ (e.g., matrix multiplication, Fourier transform, element-wise addition/multiplication over an array, etc.). Then in order for reverse-mode differentiation to work, the code needs to define how to go from the $\partial f/\partial a_j'$ to the $\partial f/\partial a_i$ during the backward pass, and it needs to store the corresponding necessary variables during the forward pass (for instance, in PyTorch this can be done for custom operations by defining the methods \texttt{forward} and \texttt{backward} of a class derived from \texttt{torch.autograd.Function}, and by using the function \texttt{save\_for\_backward} in the \texttt{forward} method). In practice, different operations may require storing different amounts of memory in the forward pass, as seen in the above example ($N^2$ scalars for step 1, but $N$ scalars for step 2).

\section{Benchmarks for the accuracy of Cheetah's space-charge implementation}
\label{app:benchmarks}

\subsection{Space charge fields in a Gaussian bunch}

In order to benchmark the space charge solver in Cheetah, we consider the calculation of the Lorentz force due to space charge within a relativistic Gaussian beam. For this benchmark, the beam has a charge $Q = 1\, \mathrm{nC}$, a transverse RMS size $\sigma_x = \sigma_y = 1\,\mathrm{mm}$, and a longitudinal RMS size $\sigma_z = 1\,\mathrm{\mu m}$. We vary the beam's Lorentz factor $\gamma$ from 10 to 10,000, and thus the aspect ratio of the beam in its rest frame $r = \sigma_z'/\sigma_x' = \gamma \sigma_z/\sigma_x$ varies from 0.01 to 10. This benchmark is therefore similar to the one performed in \cite{Mayes2018,MitchellICFA2024}.

In the beam's rest frame, the potential is given by \cite{Mayes2018}:
\[ \phi'(x', y', z') = \frac{Q}{4\pi \epsilon_0}\sqrt{\frac{2}{\pi}} \int_0^\infty \frac{e^{-\frac{\lambda^2 x'^2}{2(\lambda^2\sigma_x'^2 + 1)}-\frac{\lambda^2 y'^2}{2(\lambda^2\sigma_y'^2 + 1)}-\frac{\lambda^2 z'^2}{2(\lambda^2\sigma_z'^2 + 1)}}}{\sqrt{(\lambda^2\sigma_x'^2 + 1)(\lambda^2\sigma_y'^2 + 1)(\lambda^2\sigma_z'^2 + 1)}} d\lambda \]
where primed quantities indicate values in the rest frame. Since $\boldsymbol{B} = \boldsymbol{0}$ for the space-charge field in the beam frame, the Lorentz force is $\boldsymbol{F}'=-q\boldsymbol{\nabla}_{\boldsymbol{x}'}\phi'$. Using a Lorentz transform at fixed time $t=0$ in the laboratory frame (for which $x', y', z' = x, y, \gamma z$), we can readily find the Lorentz force in the laboratory frame:
\begin{align} 
&F_x(x,y,z) = \frac{F_x'(x', y', z')}{\gamma} \nonumber\\ 
&=\frac{q Q x}{4\pi \gamma\epsilon_0}\sqrt{\frac{2}{\pi}} \int_0^\infty \frac{\lambda^2 e^{-\frac{\lambda^2 x^2}{2(\lambda^2\sigma_x^2 + 1)}-\frac{\lambda^2 y^2}{2(\lambda^2\sigma_y^2 + 1)}-\frac{\lambda^2 \gamma^2z^2}{2(\lambda^2\gamma^2\sigma_z^2 + 1)}}}{\sqrt{(\lambda^2\sigma_x^2 + 1)^3(\lambda^2\sigma_y^2 + 1)(\lambda^2\gamma^2\sigma_z^2 + 1)}} d\lambda 
\label{eq:Fx}
\end{align}
\begin{align}
&F_z(x,y,z) = F_z'(x', y', z') \nonumber \\
&=\frac{q Q \gamma z}{4\pi \epsilon_0}\sqrt{\frac{2}{\pi}} \int_0^\infty \frac{\lambda^2 e^{-\frac{\lambda^2 x^2}{2(\lambda^2\sigma_x^2 + 1)}-\frac{\lambda^2 y^2}{2(\lambda^2\sigma_y^2 + 1)}-\frac{\lambda^2 \gamma^2z^2}{2(\lambda^2\gamma^2\sigma_z^2 + 1)}}}{\sqrt{(\lambda^2\sigma_x^2 + 1)(\lambda^2\sigma_y^2 + 1)(\lambda^2\gamma^2\sigma_z^2 + 1)^3}} d\lambda 
\label{eq:Fz}
\end{align}

\begin{figure}
    \centering
    \includegraphics[width=\linewidth]{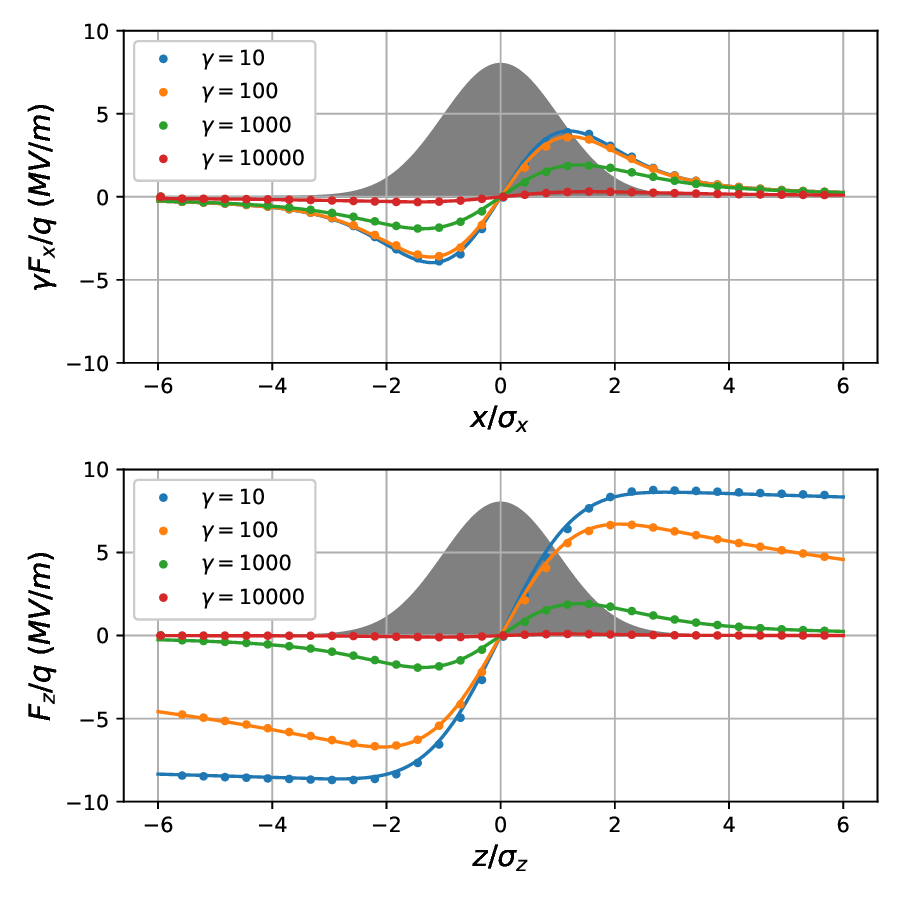}
    \caption{Plots of the Lorentz force $\boldsymbol{F} = q(\boldsymbol{E}+ \boldsymbol{v}\times\boldsymbol{B})$ felt by particles of a relativistic beam in the lab frame, due to space charge, for various values of the Lorentz factor $\gamma$. The solid lines are obtained by evaluating Eq.~(\ref{eq:Fx}) and Eq.~(\ref{eq:Fz}), while the dots are from the space charge calculation within Cheetah.}
    \label{fig:benchmark_B1}
\end{figure}

These theoretical formulas are plotted in Fig.~\ref{fig:benchmark_B1} (solid lines) and compared with the predictions of Cheetah's space charge solver (dots). The Cheetah calculation used $10^6$ macroparticles and a grid extending over $\pm 6\sigma_x$, $\pm 6\sigma_y$, $\pm 6\sigma_z$ in $x$, $y$ and $z$ respectively, with $128^3$ cells. As can be seen in Fig.~\ref{fig:benchmark_B1}, excellent agreement is found between the theoretical formulas and the Cheetah predictions.

\subsection{Free expansion of a cold uniform bunch}

We also evaluate the accuracy of Cheetah for the case of the free expansion of a cold, uniformly charged bunch. The setup is identical to that described in Sec.\ref{sec:structure} (see also \cite{MitchellICFA2024}), and we use the same numerical and physical parameters listed in Table~\ref{tab:parameters}, with the exception of $N_{\mathrm{slices}}$ which is set to $N_{\mathrm{slices}}=10$ here for increased accuracy.

With these parameters, the bunch is spherical in its rest frame, and thus the space charge forces are linear inside the bunch ($\boldsymbol{E} \propto \boldsymbol{r}$ in the rest frame). As a result, the bunch remains spherical and uniformly charged as it expands, and its radius $R$ evolves according to the following envelope equation (with no external focusing and zero emittance) \cite{Reiser}:
\begin{equation}
\frac{d^2 R}{ds^2} = \frac{N_b r_c}{\gamma^2\beta^2 R^2}
\label{eq:envelope}
\end{equation}
where $R$ is the beam radius, $N_b$ is the number of beam particles, $r_c$ is the classical electron radius, and $\gamma$ and $\beta$ are the Lorentz factor and normalized velocity of the beam, respectively.

\begin{figure}
    \centering
    \includegraphics[width=\linewidth]{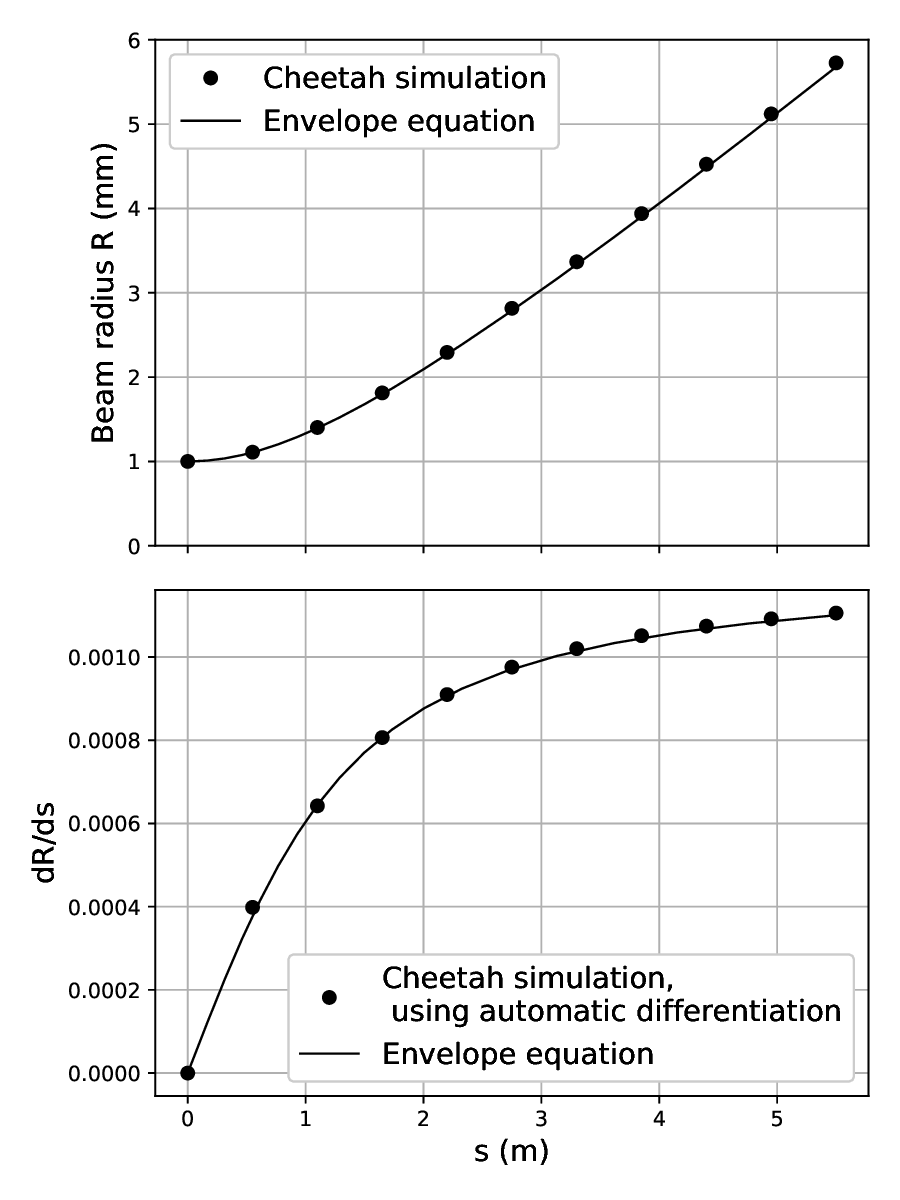}
    \caption{Comparison of beam radius evolution during the free expansion of a cold, uniformly charged bunch, as predicted by numerical integration of the envelope equation Eq.~(\ref{eq:envelope}) and by Cheetah simulations. The setup uses the parameters from Table~\ref{tab:parameters} with $N_{\mathrm{slices}}=10$.}
    \label{fig:benchmark}
\end{figure}

The top panel fo Fig.~\ref{fig:benchmark} compares the results of the numerical integration of Eq.~(\ref{eq:envelope}) with results obtained using Cheetah. In the case of Cheetah, the beam radius is determined using $R = \sqrt{5 \langle x^2 \rangle}$, which is valid for a uniformly charged, spherical beam, where $\langle .. \rangle$ denotes an average over macroparticles. In addition, in order to benchmark the accuracy of Cheetah's automatic differentiation, we also compare $d R/ds$ in the lower panel of Fig.~\ref{fig:benchmark}. We emphasize that, in the case of Cheetah, the values of $dR/ds$ on Fig.~\ref{fig:benchmark} are not obtained by an approximate numerical method (e.g., finite difference from the obtained values of $R(s)$), but instead the exact derivatives $dR/ds$ are obtained using Cheetah's built-in automatic differentiation. As can be seen in Fig.~\ref{fig:benchmark}, there is excellent agreement between the predictions of the envelope equation and those of Cheetah, for both $R(s)$ (upper panel) and $dR/ds$ (lower panel).

\section{Measuring memory usage}
\label{app:memory}
In this paper, the memory usage of Cheetah was measured using two built-in PyTorch tools for measuring memory use on GPUs. First, PyTorch's \texttt{torch.profiler.profile} was used to generate the memory graphs shown in Fig. \ref{fig:timeline}. This profiler includes the necessary traces to generate a memory graph and annotate it.

Second, \texttt{torch.cuda.memory\_allocated()} was used to determine the memory before and after each selected part of the space charge kick to measure the $\Delta M$ plotted in \cref{fig:charge_deposition,fig:coefficients,fig:field_gather,fig:green_function_calculation,fig:total}. Care was taken to ensure that the memory was recorded before and after function calls and that Python's \texttt{del} operator was used on temporary variables so that any vestigial memory allocations were freed by Python's garbage collector and not included in the data.

\section{Fitting technique}
\label{app:fit}

In order to generate the linear fits for the data in \cref{fig:charge_deposition,fig:coefficients,fig:field_gather,fig:green_function_calculation,fig:total}, we ran least squares. Ordinary least squares (OLS) minimizes the sum of the squared absolute errors of the fit at each point. Given $A$ is the feature matrix corresponding to the data, $y_i$ are the outputs, we compute the coefficient vector $x_\text{ols}$ for the OLS fit as follows:
$$ x_\text{ols} = \mathrm{argmin}_x \sum_i\left|Ax - y_i\right|^2$$
Because both the independent variables and the output vary over orders of  magnitude, OLS would heavily favor the data points at the extremes of the data set where the output has a high magnitude. This would cause the fit to essentially ignore the data points at the smaller end of the data set. To generate a fit that works in both the extremes and the smaller end of the data set, we instead minimize the sum of the squared \textit{relative} errors of the fit at each point. The coefficient vector for this weighted least squares fit, $x_\text{wls}$, can be computed as follows:
$$ x_\text{wls} = \mathrm{argmin}_x \sum_i\left|\frac{Ax - y_i}{y_i}\right|^2 = \mathrm{argmin}_x \sum_i\left|\frac{1}{y_i}Ax - 1\right|^2 $$
This can be seen as a OLS fit with each row of the feature matrix scaled by the inverse $y_i$ and the output set to a vector of ones. This weighted least squares fit was used instead of OLS in this paper.

\bibliographystyle{ACM-Reference-Format}
\bibliography{references}

\end{document}